\documentclass[12pt,a4paper]{article}

\usepackage[tbtags]{amsmath}
\usepackage{amssymb}
\usepackage{cite}
\usepackage[usenames]{color}
\usepackage{float}
\usepackage{graphicx}
\usepackage{graphics}
\usepackage[pdf]{optional}
\usepackage{tabularx}
\usepackage{times}
\usepackage{theorem}
\usepackage[T1]{url}

\opt{print}{%
\usepackage[ps2pdf,colorlinks=false,linkcolor=black, citecolor=black,%
  pagecolor=black,urlcolor=black,anchorcolor=black,pdfborder={0 0 0},%
  linktocpage=true,plainpages=false,pdfpagelabels=true,%
  bookmarksopen=true,pagebackref=false,breaklinks=true,%
  bookmarksopenlevel=1,a4paper=true,draft=false,%
  pdftitle={On the continuous spectral component of the Floquet operator%
    for a periodically kicked quantum system},%
  pdfauthor={James McCaw and B. H. J. McKellar},%
  pdfsubject={School of Physics/Uni Melbourne/Australia},%
  pdfkeywords={spectral analysis, floquet operator, time evolution,%
    periodic kick, quantum dynamics, spectrum, pure point,%
    singularly continuous, absolutely continuous, chaos, quantum chaos,%
    unitary, number theory, Weyl sum, rank N, rank 1, projection operator,%
    unitary operator, non-perturbative, cyclicity, kicked rotor,%
    kicked harmonic oscillator, chaotic dynamics}
  pdfcreator = {LaTeX with hyperref package},
  pdfproducer = {dvips, ps2pdf}]{hyperref}
}
\opt{pdf}{%
\usepackage[ps2pdf,colorlinks=true,linkcolor=darkblue, citecolor=darkgreen,%
  pagecolor=darkblue, urlcolor=blue,anchorcolor=black,pdfborder={0 0 0},%
  linktocpage=true,plainpages=false,pdfpagelabels=true,%
  bookmarksopen=true,pagebackref=true,breaklinks=true,%
  bookmarksopenlevel=1,a4paper=true,draft=false,%
  pdftitle={On the continuous spectral component of the Floquet operator%
    for a periodically kicked quantum system},%
  pdfauthor={James McCaw and B. H. J. McKellar},%
  pdfsubject={School of Physics/Uni Melbourne/Australia},%
  pdfkeywords={spectral analysis, floquet operator, time evolution,%
    periodic kick, quantum dynamics, spectrum, pure point,%
    singularly continuous, absolutely continuous, chaos, quantum chaos,%
    unitary, number theory, Weyl sum, rank N, rank 1, projection operator,%
    unitary operator, non-perturbative, cyclicity, kicked rotor,%
    kicked harmonic oscillator, chaotic dynamics}
  pdfcreator = {LaTeX with hyperref package},
  pdfproducer = {dvips, ps2pdf}]{hyperref}
}

\definecolor{darkblue}{rgb}{0,0,0.5}
\definecolor{darkgreen}{rgb}{0,0.5,0}
\definecolor{darkred}{rgb}{0.5,0,0}

\theorembodyfont{\rmfamily \itshape} \newtheorem{conj}{Conjecture}[section]
\theorembodyfont{\rmfamily \itshape} 
\theorembodyfont{\rmfamily \itshape} \newtheorem{defn}[conj]{Definition}
\theorembodyfont{\rmfamily \itshape} \newtheorem{lemma}[conj]{Lemma}
\theorembodyfont{\rmfamily \itshape} 
\theorembodyfont{\rmfamily \itshape} \newtheorem{thrm}[conj]{Theorem}
\theoremheaderfont{\bf\scshape}

\newcommand{\cotg}{\operatorname{cotg}}

\newcommand{\tr}{\operatorname{Tr}}
\newcommand{\refeq}[1]{(\ref{#1})}

\newcommand{\refsec}[1]{Section~\ref{#1}}
\newcommand{\etal}{\emph{et.\ al.}}

\newcommand{\ac}{\text{ac}}
\newcommand{\cs}{\text{sc}}
\newcommand{\cont}{\text{cont}}

\newcommand{\sing}{\text{s}}
\newcommand{\lpar}{\parallel \negthickspace}
\newcommand{\rpar}{\negthickspace \parallel}

\newcommand{\proofend}{\hfill$\Box$\newline}

\title{On the continuous spectral component of the Floquet operator for
a periodically kicked quantum system}
\author{
  James~McCaw\thanks{Electronic mail: j.mccaw@physics.unimelb.edu.au}
  \phantom{a}and B.~H.~J.~McKellar\thanks{Electronic mail: b.mckellar@physics.unimelb.edu.au} \\
  \textit{School of Physics, Research Centre for High Energy Physics,} \\
  \textit{The University of Melbourne, Victoria, 3010, Australia.}
}

\date{(Dated: \today)}

\begin{document}

\raggedbottom

\maketitle


\begin{abstract}
By a straightforward generalisation, we extend the work of Combescure
[J.\ Stat.\ Phys.\ \textbf{59}, 679 (1990)] from rank-$1$ to rank-$N$
perturbations. The
requirement for the Floquet operator to be pure point is established and
compared to that in Combescure. The result matches that in
McCaw and McKellar [J.\ Math.\ Phys.\ \textbf{46}, 032108 (2005)]. The
method here is an alternative to that work. We show
that if the condition for the Floquet operator to be pure point is relaxed,
then in the case of the $\delta$-kicked Harmonic oscillator, a singularly
continuous component of the Floquet operator spectrum exists. We also
provide an in-depth discussion of the conjecture presented in the work of
Combescure of the case where the unperturbed Hamiltonian is more
general. We link the physics conjecture directly to a number-theoretic
conjecture of Vinogradov [\textit{The Method of Trigonometrical Sums in the
Theory of Numbers} (Interscience, London, 1954)] and show that a solution of
Vinogradov's conjecture solves the physics conjecture. The result is
extended to the rank-$N$ case. The relationship between our work and the
work of Bourget [J.\ Math.\ Anal.\ Appl.\ \textbf{276}, 28 (2002);
\textbf{301}, 65 (2005)] on the physics conjecture is discussed.
\end{abstract}


\section{\label{sec:c_int}Introduction}

The spectral analysis of the Floquet operator (the unitary time-evolution
operator over a single kick period) is of great interest for periodically
perturbed Hamiltonian systems. There are general arguments
\cite{McCaw05.1,Antoniou02,Antoniou03,Combes,Milek90.1} which indicate that
an understanding or classification of the spectrum of the time-evolution
operator can provide information on the dynamics of the system. In
particular, the existence of a singularly continuous spectrum of the
Floquet operator allows for a slow diffusive energy growth over time,
typical of a chaotic system. Thus, this work has significance in the broad
field of quantum chaos. For a more detailed discussion of the links among 
spectral analysis, dynamics and chaos, see the introductory sections of
\cite{McCaw05.1} and references therein.

The work in \cite{McCaw05.1} established a non-perturbative stability
result on the spectral nature of the Floquet operator for simple systems
with a rank-$N$ perturbation periodic in time. The conditions under which
the Floquet spectrum remains pure point were established. Here, we will
first show the same result, but in a very different manner, before
proceeding to determine when a continuous spectrum may arise. This result
sheds further light on the array of possible dynamics that periodically
perturbed systems may experience.

We consider Hamiltonians of the form
\begin{equation}
\label{eq:c_hamiltonian}
  H(t) = H_0 + \left(\sum_{k=1}^N\lambda_k|\psi_k\rangle\langle\psi_k|\right)
    \sum_{n=0}^\infty \delta(t-nT)\text{,}
\end{equation}
where $\lambda_k \in \mathbb{R}$ and each vector $|\psi_k\rangle$ is a
linear combination of the $H_0$ basis states, $|\phi_n\rangle$,
\begin{equation*}
|\psi_k\rangle = \sum_{n=0}^\infty (a_k)_n | \phi_n\rangle\text{.}
\end{equation*}
The states $| \psi_k\rangle$ are orthogonal
\begin{equation*}
\langle\psi_k | \psi_l\rangle = \delta_{kl}\text{.}
\end{equation*}
The Floquet operator\footnote{Our Floquet operator differs from that in
Combescure \cite{Combescure90} and Bourget \cite{Bourget02.1,Bourget05}. An
erronous $T$ was introduced in \cite{Combescure90} and it has been carried
through in the literature. Note that the theorems proved therein
are not invalidated in any way by this error.} is
\begin{equation*}
V \equiv U(T) = e^{-iH_0T/\hbar}
e^{-i(\sum_k \lambda_k |\psi_k\rangle\langle\psi_k|)/\hbar}\text{.}
\end{equation*}

The basic result, as established in \cite{McCaw05.1} is that if
every $|\psi_k\rangle$ is in $l_1(H_0)$, the
spectrum will remain pure point for almost every perturbation strength.

If this condition is dropped for any one of the $|\psi_k\rangle$, then we
no longer have $V_{\lambda_1,\ldots,\lambda_N}$ pure point. In fact, on the
subspace $\mathcal{H}_k$ defined by that space for which $|\psi_k\rangle$ is
a cyclic vector for operator $U$, the spectrum is purely continuous. At this
point, we note that Milek and Seba \cite{Milek90.1} have incorrectly
concluded from Combescure's work that the existence of a $\psi$ such that
$\psi$ is in the continuous subspace of $\mathcal{H}$ implies that the
whole of $\mathcal{H}$ is continuous for the operator $V$. This statement
would require the assumption that $\psi$ is a cyclic vector for $U$, which
is simply impossible for $|\psi\rangle\langle\psi|$ as an arbitrary
projection.

For Milek and Seba's work to be properly justified, we show that a
sufficient condition is that Vinogradov's number-theoretic conjecture,
stated over fifty year ago,\footnote{The reference is to the 1954 English
translation of Vinogradov's original work, published in 1947. The work in
Vinogradov's 1947 monograph incorporates results from a series of papers
and a first monograph from 1937. It is unknown to us when the
conjecture referred to was first presented, but it was at least fifty years
ago.} is true. This observation is linked to the conjecture put forward by
Combescure \cite{Combescure90} and partially addressed by Bourget
\cite{Bourget02.1}. After the completion of this work, we became aware of
a recent paper by Bourget \cite{Bourget05} which successfully resolves the
issues with Milek and Seba's work by building on the earlier work in
\cite{Bourget02.1}. Bourget's new work in no way invalidates
the arguments presented here---the two approaches are
complimentary.

In \refsec{sec:c_comb1} we extend Combescure's rank-$1$ theorem on the pure
point spectral nature of $V$ to the rank-$N$ case. In \refsec{sec:c_comb2}
we then show the existence of a continuous spectrum for the case where $H_0$
is the harmonic oscillator and the perturbation is rank-$N$. In
\refsec{sec:c_genspec} we investigate Combescure's conjecture, the 
answer provided by Bourget and the link to number theory and Vinogradov's
conjecture. Finally, in \refsec{sec:c_milek}, we extend Milek and Seba's work
to the rank-$N$ case, correcting a number of subtle errors. We emphasise
that their work has only recently been fully justified (by Bourget in
\cite{Bourget05}). We provide a complimentary justification, linked to the
number-theoretic investigations and Vinogradov's conjecture just mentioned.


\section{\label{sec:c_comb1}Rank-N generalisation of Combescure's first
theorem}

We consider the measures
\begin{equation*}
m_{k,\lambda_k} = \langle\psi_k | E_{\lambda_k}(S) | \psi_k\rangle\text{.}
\end{equation*}
Each $|\psi_k\rangle$ admits a cyclic subspace of $\mathcal{H}$,
$\mathcal{H}_k$. As argued in the later part of the proof of
Theorem~4.3 in \cite{McCaw05.1}, on the space
$\mathcal{H}\ominus \left(\bigoplus_{k=1}^N \mathcal{H}_k\right)$, the
perturbation
\begin{equation*}
\sum_{k=1}^N \lambda_k |\psi_k\rangle\langle\psi_k|
\end{equation*}
is null and thus $V=U$ is trivially pure point. Henceforth, we may safely
restrict our proof to the subspace $\bigoplus_{k=1}^N \mathcal{H}_k$ for
which the vectors $|\psi_k\rangle$ form a cyclic set.

Directly following Combescure, the measure for a point
$x\in[0,2\pi)$ for the operator $V$ acting on the state $|\psi_k\rangle$
is given by
\begin{equation}
\label{eq:c_measure}
m_{k,\lambda_k}(\{x\}) = \frac{-4(1+\mu_k)}{\mu_k^2}B_k(x)\text{,}
\end{equation}
where
\begin{equation*}
\mu_k = e^{i\lambda_k/\hbar} - 1
\end{equation*}
and
\begin{equation*}
B_k(x) = \left[\int_0^{2\pi}
dm_{k,\lambda_k=0}(\theta)\left(\sin^2\left[(x-\theta)/2\right]\right)^{-1}
  \right]^{-1}\text{.}
\end{equation*}
This result is the essence of Lemma~1 in \cite{Combescure90}. When
$H_0$ is pure point, it is a trivial calculation to show that
\begin{equation}
\label{eq:c_Binv}
B_k^{-1}(x) = \sum_{n=0}^\infty \frac{|(a_k)_n|^2}
  {\sin^2\left[(x-\theta_n)/2\right]}\text{.}
\end{equation}
Corollary~2 in \cite{Combescure90} is replaced with the following.

\begin{thrm}
\label{thrm:c_comb1}
Assume $H_0$ is pure point, with $\{\phi_n\}_{n\in\mathbb{N}}$ and
$\{\alpha_n\}_{n\in\mathbb{N}}$ as eigenstates and eigenvalues. Let each
\begin{equation*}
\psi_k = \sum_{n=0}^\infty (a_k)_n\phi_n
\end{equation*}
be cyclic for $H_0$ (hence, cyclic for $U$ and $V$) on $\mathcal{H}_k$ and
$\langle\psi_k | \psi_l\rangle = \delta_{kl}$. Then $e^{ix}$ belongs to the
point spectrum of $V_{\lambda_1,\ldots,\lambda_N}$ if and only if
\begin{equation*}
\prod_{k=1}^N B_k^{-1}(x) < \infty\text{,}
\end{equation*}
where
\begin{equation*}
\theta_n = 2\pi\{\alpha_n/2\pi\hbar\}\text{,}
\end{equation*}
$\{z\}$ being the fractional part of $z$.
\end{thrm}

\emph{Proof.} (\ref{thrm:c_comb1})
The proof follows that in \cite{Combescure90}. By the cyclicity of each
$|\psi_k\rangle$ on $\mathcal{H}_k$ and the argument in Theorem~4.3 of
\cite{McCaw05.1}, $e^{ix}$ is an eigenvalue of $V_{\lambda_1,\ldots,\lambda_N}$
if and only if every $m_{k,\lambda_k}(\{\theta\}) \neq 0$ at $\theta=x$. As
already mentioned, using
\begin{equation*}
dm_{k,\lambda_k=0} = \sum_{n=0}^\infty |(a_k)_n|^2
  \delta(\theta-\theta_n)d\theta
\end{equation*}
we obtain, for each $k$,
\begin{equation*}
B_k^{-1}(x) = \sum_{n=0}^\infty \frac{|(a_k)_n|^2}{\sin^2[(x-\theta_n)/2]}
\text{.}
\end{equation*}
We now consider the eigenvalue $e^{ix}$. If it were to be that for some
$k$, $m_{k,\lambda_k}(\{x\}) = 0$, then we would have found a vector, namely
$|\psi_k\rangle$, such that $V|\psi_k\rangle$ was continuous. We have in
fact found that the whole subspace $\mathcal{H}_k$ is continuous.
Thus, for $V$ to be pure point, we require every
$m_{k,\lambda_k}(\{\theta\})\neq 0$. Thus, we are lead to consider the
requirement
\begin{equation*}
\prod_{k=1}^N B_k^{-1}(x) < \infty\text{.}
\end{equation*}
\proofend

As in \cite{Combescure90}, the relationship
\begin{equation}
\label{eq:c_cotg}
\sum_{n=0}^\infty |(a_k)_n|^2\cotg\left(\frac{x-\theta_n}{2}\right)
  = \cotg\frac{\lambda_k}{2\hbar}
\end{equation}
also holds for each $k$. To show \refeq{eq:c_cotg}, we consider each $k$
separately. The proof is the same as for the rank-$1$ case.
See \cite{Combescure90}. Points to consider are that each projection
operator in the rank-$N$ projection is normalised and hence for every $k$
we have
\begin{equation*}
\sum_{n=0}^N |(a_k)_n|^2 = 1\text{.}
\end{equation*}

\proofend

In order to complete the generalisation of Combescure's first theorem, we
require, just as in \cite{Combescure90}, two additional Lemmas.

\begin{lemma}
If $\sum_{n=0}^\infty |(a_k)_n| < \infty$, then $B_k^{-1}(x)<\infty$ for
almost every $x\in\mathbb{R}$.
\end{lemma}

For each $k\in 1,\ldots,N$, the proof is identical to that in
\cite{Combescure90}.

\begin{lemma}
The following two statements are equivalent.
\begin{enumerate}
\item For almost every $(\lambda_1,\ldots,\lambda_N)$,
  $V_{\lambda_1,\ldots,\lambda_N}$ has only a point spectrum.
\item For every $k\in\{1,\ldots,N\}$ and for almost every $x$,
$B_k(x)\neq 0$.
\end{enumerate}
\end{lemma}

The proof is again virtually identical to Combescure's proof. For each $k$,
the continuous part of the spectrum is supported outside the set
$E_k = \{x\in[0,2\pi): B_k(x)\neq 0\}$ and, for $\lambda_k\neq 0$,
the point part of $dm_{k,\lambda_k}$ is supported by the set $E_k$. Thus,
for $V_{\lambda_1,\ldots,\lambda_N}$ to be pure point for almost every
$\lambda_1,\ldots,\lambda_N$ and for every $k$, we require
\begin{equation*}
m_{k,\lambda_k}([0,2\pi)\setminus E_k) = 0\text{.}
\end{equation*}
This in turn implies that for every $k$
\begin{equation*}
\int_0^{2\pi} d\lambda_k'h(\lambda_k')m_{k,\lambda_k}([0,2\pi)\setminus E_k)
  = 0\text{,}
\end{equation*}
where $\lambda_k' = \lambda_k/\hbar$ and
\begin{equation*}
h(\lambda) = 2\Re\frac{1}{1-ce^{i\lambda}}
\end{equation*}
for some $|c|<1$.

Lemma 5 in \cite{Combescure90} trivially applies for each $k$. Thus, we have
generalised Combescure's work to obtain the result that the Floquet
operator for the rank-$N$ perturbed Hamiltonian has a pure point spectrum.
The result matches that in \cite{McCaw05.1}.


\section{\label{sec:c_comb2}Rank-N generalisation of Combescure's second
theorem}

Having shown that the Floquet operator remains pure point for perturbations
constructed from the vectors $|\psi_k\rangle\in l_1(H_0)$, Combescure
relaxes this condition to allow for the emergence of a continuous spectral
component of the Floquet operator. This result is easily generalised to the
rank-$N$ case. The key point is that the technique in \cite{Combescure90}
applies independently for each $k$. We do not discuss the details of the
rank-$1$ proof here at all, delaying an analysis to \refsec{sec:c_genspec}
where we will have the opportunity to generalise the result still further.
Here, we simply provide the argument for why each $k$ may be treated
independently.  Before proceeding, some subtleties of what Combescure
actually shows are highlighted. They are seemingly overlooked by some in
the literature (e.g.,\ \cite{Milek90.1}).

The cyclicity requirement was essential in the proof that the Floquet
operator spectrum was pure point. Here, we can happily ignore the
cyclicity conditions, as our only goal is to establish the
existence of a state in the continuous subspace $\mathcal{H}_\cont$.
We need not try and ensure the result obtained by
considering $\langle\psi|E(S)|\psi\rangle$ is applicable to all other
vectors in $\mathcal{H}$---the very idea is ill-formed as the perturbation
is null on a subset of $\mathcal{H}$ and thus there is always part of
$\mathcal{H}$ where $V$ has a discrete spectrum. Milek and Seba seem to
have missed this point, restating Theorem~1 in \cite{Combescure90} in a
way that implies that all $\psi$ are in $\mathcal{H}_\cont$.

If
\begin{equation*}
\langle\psi_k|E(\{x\})|\psi_k\rangle = 0 
\end{equation*}
then $\mathcal{H}_{ac}$ contains at least the state
$|\psi_k\rangle$. The point to be mindful of is that this does not allow
one to conclude that the Hilbert space for the operator $V$ has
$\mathcal{H}_{pp} = \emptyset$, as implied by Milek and Seba
\cite{Milek90.1}. To draw that conclusion would require an
argument to show that a cyclic vector does in fact exist for $V$. This does
not seem possible in the general context we have here.

Combescure's proof (Lemma~6 in \cite{Combescure90}) that $\sigma_\cont(V)
\neq \emptyset$ is based on showing
that $B^{-1}(x)\rightarrow\infty$ (equation \refeq{eq:c_Binv}). As the
spectral measure of a single point $x$ is proportional to $B(x)$
(equation \refeq{eq:c_measure}), if $B^{-1}(x)\rightarrow\infty$, then the
contribution of the single point is zero. That is, $e^{ix}$ is in the
continuous spectrum of the Floquet operator. Combescure argues
(see \refsec{sec:c_genspec} for details) that
\begin{equation*}
B^{-1}(x) \geq \#S(x)
\end{equation*}
where $\#S(x)$ is the number of element of a particular set $S$. She then
shows (the bulk of the proof) that $\#S(x)\rightarrow\infty$ and thus
$B^{-1}(x)\rightarrow\infty$. We generalise the result in a straightforward
manner.

\begin{thrm}
\label{thrm:c_comb2}
Assume $\alpha_n = n\hbar\omega$ with $\omega$ irrational.
If $|\psi_k\rangle \notin l_1(H_0)$ for at least one $k\in 1,\ldots,N$,
then $\sigma_\cont(V) \neq \emptyset$.
\end{thrm}

\emph{Proof.} (\ref{thrm:c_comb2})
Following the same argument as for the rank-$1$ case, we take
\begin{equation*}
|(a_k)_n| = n^{-\gamma}2\pi
\end{equation*}
for the state $|\psi_k\rangle$, in such a way that the condition
$\langle\psi_k|\psi_l\rangle=\delta_{kl}$ is preserved.

With this construction, the proof that the number of elements in $S(x)$ is
infinite \cite{Combescure90} applies to each subsequence $S_k(x)$. The
number of elements, $\#S_k(x)$, in each sub-sequence for which
$|psi_k\rangle \notin l_1(H_0)$, is infinite. The Floquet operator for the
rank-$N$ perturbed harmonic oscillator obtains a continuous spectral
component.
\proofend

\subsection{Discussion}

It must be noted that the proof of Lemma~6 in \cite{Combescure90} is only
valid for the eigenvalue spectrum,
\begin{equation*}
\alpha_n = n\hbar\omega\text{,}
\end{equation*}
of the harmonic oscillator. Combescure does however conjecture that the
argument will be valid for more general eigenvalue spectra, including the
rotor
\begin{equation*}
\alpha_n \propto n^2\text{.}
\end{equation*}
For Milek and Seba's numerical work (using the rotor) to be based on
valid mathematical arguments, a proof of this conjecture is required.
It was only very recently that a proof was developed \cite{Bourget05}, some
fifteen years after the numerical results of Milek and Seba were published.
As already mentioned, the work presented here, aiming to justify Milek and
Seba's numerical work, was developed independently and is complimentary to
Bourget's approach. 

In \refsec{sec:c_genspec} we show that if
a conjecture from number theory on the estimation of exponential sums is
true, then Milek and Seba's work can be justified. The rank-$N$
generalisation is straightforward. Considering the number theory
conjecture has stood for some fifty years, it seems we may have to wait 
quite some time for a proof.\footnote{Of course, now that Bourget has
provided a direct proof for the rotor, the numerical simulations are on
safe ground.}

For more general eigenvalue spectra (loosely $\alpha_n \propto n^j$) the
situation is similar. For $j\geq 3$ Bourget \cite{Bourget02.1} made
significant progress. He has now covered the $j=2$ case \cite{Bourget05}. A
continuous component of the Floquet operator exists for certain
constructions of $|\psi\rangle$. The conditions, for all $j\geq 2$, are
complicated and more restrictive than the $|\psi\rangle \notin l_1(H_0)$
condition for the harmonic oscillator.  The result is easily extended to
the rank-$N$ case due to the independence of each $k$ as already discussed.
Here, by utilising a number-theoretic conjecture, we will provide
improvements to the work of Bourget (both the $j\geq 3$ and $j=2$ cases).
See \refsec{sec:c_genspec}.

Returning to the harmonic oscillator case, by applying
Theorem~\ref{thrm:c_comb2} we may conclude that for each
$|\psi_k\rangle \notin l_1(H_0)$, $\mathcal{H}_k$ is purely continuous.
Thus, by dropping the $l_1$ condition for all $|\psi_k\rangle$, we have
shown that $V$ is purely continuous on the subspace of $\mathcal{H}$ where
the perturbation is non-zero. On the subspace of $\mathcal{H}$ where the
perturbation is zero, $V=U$ trivially and thus that portion of the
Hilbert space remains pure point.


\section{\label{sec:c_genspec}Discussion of Combescure's conjecture}

Combescure \cite{Combescure90} makes a remark (Remark c.)\ that she
believes Theorem~\ref{thrm:c_comb2}
(Lemma~6, \cite{Combescure90}) is generalisable to include
systems other than the harmonic oscillator. Explicitly, she conjectures
that Hamiltonians, $H_0$, with eigenvalues, $\alpha_n$, of the form
\begin{equation}
\label{eq:c_eigenvalues}
\alpha_n = \hbar \sum_{j=0}^p \beta_j n^j
\end{equation}
with $\beta_j T / 2\pi$ Diophantine for some $j: 1 \leq j \leq p$ will
have the vector $\psi$ in the continuous spectral subspace of $V_\lambda$.

At an intuitive level, one would expect this to be true. The precise nature
of the eigenvalue spectra (proportional to $n$ or a polynomial in $n$)
should not make a significant difference. Milek \cite{Milek89}
argues that Combescure's work can be used in the $n^2$ case based on
evidence from some numerical work that shows that the sequences obtained
are ``almost random''---however, the argument is not entirely convincing to
us. The cited numerical work of Casati \etal\
\cite{Casati85} discusses the existence of
correlations in the energy levels, rather than the lack of correlations.
While the deviations from a Poisson distribution look small to the naked
eye, Casati \etal\ \cite{Casati85} find deviations from the expected
Poisson distribution of up to 17 standard deviations. The energy levels are
correlated---it is arguable that they are not characterisable as ``almost
random'' as Milek asserts.

In late 2002, Bourget \cite{Bourget02.1} produced a proof of a slightly
modified conjecture for all but the $p=2$ case in \refeq{eq:c_eigenvalues}.
The techniques used by Bourget are similar to those followed in this work.
We will analyse Bourget's work, and highlight the key breakthrough made.
We also provide a modified argument to obtain the proof which is, we
believe, significantly easier to follow. Importantly, it also covers the
$p=2$ case, unresolved by Bourget (until very recently) due to technical
difficulties. However, it
comes at the expense of relying upon a conjecture. Our result plays a
complementary role in understanding, or perhaps appreciating, Bourget's
proof. The reliance on the conjecture removes the need for much of
the technical wizardry in Bourget's proof, and also
strengthens the work. Our analysis also indicates, or highlights, that
Combescure's conjecture is solved by a number-theoretic conjecture that has
stood for over fifty years. What seems a perfectly reasonable conjecture on
physical grounds is shown to be directly related to an abstract
mathematical conjecture.

In what follows, we rely heavily upon the lemmas and theorems in Chapter~2
of \cite{Kuipers}. We also use some results on Weyl sums from
\cite{Vinogradov}. Of key importance is an understanding of the proof of
Lemma~6 in \cite{Combescure90} on the emergence of a continuous spectrum for
the kicked harmonic oscillator. This will be discussed at the appropriate time
in this section.

\subsection{Number theory}

To discuss the conjecture, we require two concepts from
number theory---the classification of irrational numbers and the
\emph{discrepancy} of a sequence. We first introduce the concepts and
define the relevant ideas. We then proceed to analyse the conjecture
and the proof provided by Bourget. As the discussion progresses, the new
work that we have done will be presented.

For any number $\beta$, we define
\begin{itemize}
\item $[\beta]$, the integer part of $\beta$,
\item $\{\beta\}$, the fractional part of $\beta$, and
\item $\langle \beta \rangle = \min(\{\beta\}, 1-\{\beta\})$.
\end{itemize}
$\langle\beta\rangle$ is simply the ``distance to the nearest integer''.
Definition~\ref{defn:c_type_eta} is taken directly from Kuipers and
Niederreiter (Definition~3.4, p.~121, \cite{Kuipers}).

\begin{defn}
\label{defn:c_type_eta}
Let $\eta$ be a positive real number or infinity. The irrational $\beta$
is of type $\eta$ if $\eta$ is the supremum of all $\tau$ for which
\begin{equation}
\label{eq:c_type_eta}
\varliminf_{n\rightarrow\infty} q^\tau \langle q\beta \rangle = 0\text{,}
\end{equation}
where $q$ runs through the positive integers.
\end{defn}

The idea behind this definition can be seen by considering \emph{rational}
$\beta = p/q'$ for some $p$ and $q'$. Run through the positive
integers $q$. At $q=q'$, $\langle q\beta \rangle= 0$, so there is no
supremum $\eta$ for $\tau$ in \refeq{eq:c_type_eta}. In effect,
$\eta\rightarrow\infty$. For irrational $\beta$,
$\langle q\beta \rangle$ is never equal to zero but will approach zero. If
the approach is very slow, then a small $\tau$ is enough to prevent
\refeq{eq:c_type_eta} from approaching zero. $\langle q\beta \rangle$
approaching zero slowly is, in a sense, indicative of $\beta$ being badly
approximated by rational numbers. Even for very large $q'$, $p/q'$ remains
a poor approximation to $\beta$. Thus, the smaller $\eta$, the stronger
the irrationality of $\beta$. This is reasonable in the sense that rational
$\beta$s act like numbers with $\eta\rightarrow\infty$. As stated in
\cite{Kuipers}, all numbers $\beta$ have type $\eta\geq 1$.

We now define the discrepancy of a sequence---a measure of the
non-uniformity of the sequence. We consider a sequence of
numbers\footnote{Equivalently, consider any sequence $x_n$ and consider the
discrepancy of the sequence modulo~1.} $x_n$ in $[0,1)$
\begin{equation*}
\omega = (x_n)_{n\in\mathbb{N}} \text{ with } x_n\in[0,1)\text{.}
\end{equation*}
For $0\leq a < b \leq 1$ and positive integer $N$, $A([a,b),N)$ counts the
number of terms of the sequence (up to $x_N$) contained in the interval
$[a,b)$,
\begin{equation*}
A([a,b),N) = \#\{n \leq N: x_n\in[a,b)\}\text{.}
\end{equation*}

\begin{defn}
\label{defn:c_discrepancy}
The discrepancy $D_N$ of the sequence $\omega$ is
\begin{equation}
\label{eq:c_discrepancy}
D_N(\omega) = \sup_{0\leq a < b \leq 1}
  \left| \frac{A([a,b),N)}{N} - (b-a)\right|\text{.}
\end{equation}
\end{defn}

If the sequence $\omega$ is uniformly distributed in $[0,1)$ then
$D_N\rightarrow 0$ as $N\rightarrow\infty$. In this case, every interval
$[a,b]$ in $[0,1)$ gets its  ``fair share'' of terms from the sequence
$\omega$.

Estimating the discrepancy of a sequence will turn out to be vital in the
analysis of Combescure's work. The sequence of interest is basically the
eigenvalue sequence for $H_0$, but we will discuss this in greater detail
later.

The starting point for the estimations that we require is (equation~(2.42), 
Chapter~2, \cite{Kuipers}). This is a famous result
obtained by Erd\"os and Tur\'an. It states that
\begin{equation}
\label{eq:c_gendisc}
D_N \leq C\left(\frac{1}{m} +
  \sum_{h=1}^m \frac{1}{h}\left|\frac{1}{N}
  \sum_{n=1}^N e^{2\pi i h x_n} \right| \right)
\end{equation}
for any real numbers $x_1 \ldots x_N$ and any positive integer $m$.
The sum
\begin{equation*}
S = \sum_{n=1}^N e^{2\pi i h x_n}
\end{equation*}
is an example of a class of exponential sums known as \emph{Weyl sums},
reflecting the pioneering work of Weyl on providing estimations for them.
Vinogradov \cite{Vinogradov} improved on some of the estimations of Weyl.
Weyl and Vinogradov's results concern the modulus of the sum, $|S|$, and
characterise it as
\begin{equation*}
|S| \leq \gamma N\text{,}
\end{equation*}
where $N$ is the number of terms in the sum and $\gamma$ tends to zero as
$N\rightarrow\infty$. The subtle behaviour of $\gamma$ is linked to the
rational/irrational nature of the terms in the sequence.

We will use some basic results from the introductory chapter of
\cite{Vinogradov}. In general, we write
\begin{equation*}
S = \sum_{n=1}^N \exp\left(2\pi i F(n) \right)
\end{equation*}
for some function $F(n)$. The application here is when
\begin{equation*}
F(n) = \beta n^j\text{.}
\end{equation*}
For $\beta$ rational (\emph{not} the case we will be interested in)
L. K. Hau proved that $|S|$ was of order
\begin{equation*}
N^{1-(1/j) + \epsilon}
\end{equation*}
(page 3, \cite{Vinogradov}) and that this estimate could not be much
improved. Here, we are interested in the case where $\beta$ is irrational.
Estimations are much more difficult, and form the major aspect of the work
by Vinogradov. The estimations depend upon making a
rational approximation to $\beta$ and are complicated
functions of $N$ and $j$. Very loosely, he obtains results like
\begin{equation*}
|S| = O(N^{1-\rho'})
\end{equation*}
where
\begin{equation}
\label{eq:c_rho}
\rho' = \frac{1}{3(j-1)^2\log 12j(j-1)}\text{.}
\end{equation}
Vinogradov states
\begin{quote}
It is a plausible conjecture that the estimate in (\ref{eq:c_rho}) holds
with $\rho'$ replaced be $1/j - \epsilon$ \ldots\. A proof or disproof of
this conjecture would be very desirable.
\end{quote}
As the conjecture plays a central role in what follows, we state it formally.

\begin{conj}
\label{conj:c_weyl}
Consider the sum
\begin{equation*}
S = \sum_{n=1}^N \exp{2\pi i n^j\beta_j}\text{.}
\end{equation*}
For all $N$ greater than some critical value, we have
\begin{equation*}
|S| \leq cN^{1 - (1/j) + \epsilon}
\end{equation*}
for all $\epsilon > 0$ and some constant $c\in\mathbb{R}$.
\end{conj}
We do not attempt to prove Conjecture~\ref{conj:c_weyl}. Given the lengths
gone to by Vinogradov to obtain the results presented above, it seems
rather unlikely that a proof or disproof will be found any time
soon.\footnote{Incremental improvements on the estimations presented by
Vinogradov in \cite{Vinogradov} have been made over time. While Bourget
\cite{Bourget02.1,Bourget05} makes use of these improved results, the
conjecture itself remains unproven which is the only result of any
consequence in this discussion.}

\subsection{Upper and lower bounds on discrepancy}

Armed with the estimations on Weyl sums, we now proceed to derive both
upper and lower bounds on the discrepancy for sequences of the type
\begin{equation*}
\omega_j = (n^j\beta)
\end{equation*}
for $\beta$ of any type $\eta\geq 1$. It must be remembered that the upper
bound obtained is contingent upon Conjecture~\ref{conj:c_weyl}. The
lower bound obtained is not dependent upon any unproved conjectures. The
result obtained highlights the ``best possible'' nature of the conjectured
upper bound.

Firstly, (Lemma~3.2, p.~122, \cite{Kuipers}) is generalised to arbitrary
$j$.

\begin{lemma}
\label{lemma:c_3.2}
The discrepancy $D_N(\omega_j)$ of $\omega_j = (n^j\beta)$ satisfies
\begin{equation*}
D_N(\omega_j) \leq C\left(\frac{1}{m} + N^{1-(1/j)+\epsilon}
  c'\sum_{h=1}^m \frac{1}{h\langle h\beta \rangle}\right)
\end{equation*}
for any positive integer $m$ and $\epsilon>0$, where $C$ and $c'$ are
absolute constants.
\end{lemma}
\emph{Proof.} (\ref{lemma:c_3.2}) Consider \refeq{eq:c_gendisc}. It is
applicable to the first $N$ terms of the sequence $\omega_j$.
We have
\begin{equation}
\label{eq:c_242}
D_N(\omega_j) \leq C\left(\frac{1}{m} + \frac{1}{N}
  \sum_{h=1}^m \frac{1}{h}\left|
  \sum_{n=1}^N e^{2\pi i h n^j \beta} \right| \right)
\end{equation}
for any positive integer $m$. Consider the sum over $n$,
\begin{equation*}
\left| \sum_{n=1}^N e^{2\pi i h n^j \beta} \right|\text{.}
\end{equation*}
Conjecture~\ref{conj:c_weyl} allows this sum to be bounded by
\begin{equation*}
cN^{1-(1/j)+\epsilon}\text{.}
\end{equation*}
We are free to write
\begin{equation*}
c = \frac{c'}{|\sin \pi h \beta|}
\end{equation*}
as $\sin \pi h \beta$ is just some positive real number. Substituting
this result into \refeq{eq:c_242}, we obtain
\begin{equation*}
D_N(\omega_j) \leq C\left(\frac{1}{m} + N^{-(1/j) + \epsilon}
c'\sum_{h=1}^m \frac{1}{h} \frac{1}{|\sin \pi h \beta|}\right)\text{.}
\end{equation*}
Now following the argument at the end of (Lemma~3.2, \cite{Kuipers}) the
desired result is obtained.
\proofend

We now give the generalisation of (Theorem~3.2, \cite{Kuipers}). It
provides the ``best'' upper bound one could hope for when estimating
the discrepancy of the sequence $\omega_j = (n^j\beta)$. Again, remember
that the proof relies on Conjecture~\ref{conj:c_weyl}.

\begin{thrm}
\label{thrm:c_d_ub}
Assume Conjecture~\ref{conj:c_weyl} is true. Let $\beta$ be of finite
type $\eta$. Let $j$ be a positive integer $j\geq 1$. Then, for every
$\epsilon>0$, the discrepancy $D_N(\omega_j)$ of $\omega_j = (n^j\beta)$
satisfies
\begin{equation*}
D_N(\omega_j) = O\left(N^{-1/(\eta j) + \epsilon} \right)\text{.}
\end{equation*}
\end{thrm}
\emph{Proof.} (\ref{thrm:c_d_ub}) Let $\epsilon>0$ be fixed. By (Lemma~3.1 
and Lemma~3.3, p.~121--3, \cite{Kuipers}),
\begin{equation*}
\sum_{h=1}^m \frac{1}{h\langle h\beta \rangle}
  = O\left(m^{\eta-1+\epsilon'}\right)
\end{equation*}
for a fixed $\epsilon'>0$. Combining this with Lemma~\ref{lemma:c_3.2},
gives
\begin{equation*}
D_N(\omega_j) \leq C\left(\frac{1}{m}
  + N^{-(1/j)+\epsilon''}m^{\eta-1+\epsilon'}\right)
\end{equation*}
for all $m\geq 1$. Now choose $m=\left[N^{1/(\eta j)}\right]$. We obtain
\begin{equation*}
\begin{split}
D_N(\omega_j) &\leq C\left(N^{-1/(\eta j)}
  + N^{-(1/j) + \epsilon'' + (1/j) - 1/(\eta j) + \epsilon'/(\eta j)}\right) \\
  &= O\left(N^{-1/(\eta j) + \epsilon}\right)
\end{split}
\end{equation*}
where $\epsilon = \epsilon'' + \epsilon'/(\eta j)$.
\proofend

Theorem~\ref{thrm:c_d_ub} is, in a sense, optimal. For functions $f,g$,
define $f = \Omega(g)$ if $f/g\nrightarrow 0$.
\begin{thrm}
\label{thrm:c_d_lb}
Let $\beta$ be of finite type $\eta$. Let $j$ be a positive integer
$j\geq 1$. Then, for every $\epsilon>0$, the discrepancy
$D_N(\omega_j)$ of $\omega_j = (n^j\beta)$ satisfies
\begin{equation*}
D_N(\omega_j) = \Omega\left(N^{-1/(\eta j) - \epsilon} \right)\text{.}
\end{equation*}
\end{thrm}

\emph{Proof.} (\ref{thrm:c_d_lb})
Let $\epsilon>0$ be fixed.
For any given $\epsilon'>0$, there exists $0<\delta<\eta$ with
$1/(\eta-\delta)=(1/\eta)+\epsilon'$. By (Definition~3.4, p.~121,
\cite{Kuipers}), we have $\varliminf_{q\rightarrow\infty}
q^{\eta-(\delta/2)}\langle q\beta-j\rangle=0$ and thus
\begin{equation*}
\langle q\beta_j \rangle < q^{-\eta+(\delta/2)}
\end{equation*}
for an infinite number of positive integers $q$. There are infinitely
many positive integers $q$ and $p$ such that
\begin{equation*}
\left| \beta - p/q \right| < q^{-1-\eta + (\delta/2)}\text{.}
\end{equation*}
That is, by choosing $q$ large enough, we can always find a $p$ such that
$|q\beta - p| = \langle q\beta \rangle$. As $q$ increases $p/q$ is a
better approximation to the irrational $\beta$. For $\theta$ some
irrational with $|\theta|<1$, we have
\begin{equation*}
\beta = p/q + \theta q^{-1-\eta+(\delta/2)}\text{.}
\end{equation*}
Pick a $q$ such that the above relations are valid. Set
\begin{equation*}
N = \left[q^{j(\eta-\delta)}\right]\text{.}
\end{equation*}
Then for $1\leq n^j \leq N^{1/j}$,
\begin{equation*}
n^j \beta = n^j(p/q) + \theta_n\text{,}
\end{equation*}
with
\begin{equation*}
\begin{split}
\left|\theta_n\right| &=  \left|n^j\theta q^{-1-\eta+(\delta/2)}\right| \\
  &< n^j q^{-1-\eta+(\delta/2)} \\
  &\leq q^{[j(\eta-\delta)]^{1/j}-1-\eta+(\delta/2)} \\
  &=q^{-1-(\delta/2)}\text{.}
\end{split}
\end{equation*}
Thus, none of the fractional parts $\{\beta\}$, $\{2^j\beta\}$,
\ldots, $\{\left[N^{1/j}\right]\beta\}$ lie in the interval
$J=\left[q^{-1-(\delta/2)},q^{-1}-q^{-1-(\delta/2)}\right)$, so
\begin{equation*}
D_N(\omega_j) \geq \left|\frac{A(J,N)}{N}-\lambda(J)\right|
  = \lambda(J)
\end{equation*}
where $\lambda(J)$ is simply the ``size'' of the set $J$.
For large enough $q$ we have $\lambda(J)\geq 1/2q$. But from the definition
of $N$ it is clear that
\begin{equation*}
N \leq q^{j(\eta-\delta)} \leq N + 1 \leq 2N\text{,}
\end{equation*}
so
\begin{equation*}
q^{-1} \geq cN^{-\left[j(\eta-\delta)\right]^{-1}}\text{.} 
\end{equation*}
Combining these inequalities, we obtain
\begin{equation*}
\begin{split}
D_N(\omega_j) &\geq c'N^{-\left[j(\eta-\delta)\right]^{-1}} \\
 &= c'N^{-(1/j)\left(1/(\eta-\delta)\right)} \\
 &= c'N^{-(1/j)\left((1/\eta)+\epsilon'\right)} \\
 &= c'N^{-1/(\eta j) - \epsilon}
\end{split}
\end{equation*}
where $\epsilon = \epsilon'/j$. That is, we have shown, for all
$\epsilon>0$, that
\begin{equation*}
D_N(\omega_j) = \Omega\left(N^{-1/(\eta j) -\epsilon}\right)\text{.}
\end{equation*}
\proofend

\subsection{Combescure's conjecture, Bourget's work and new results}

Before discussing the conjecture, we must clearly understand Combescure's
proof for the harmonic oscillator case. As stated in \refsec{sec:c_comb2},
the aim is to show that
\begin{equation*}
B^{-1}(x) = \sum_{n=0}^\infty |a_n|^2
  \left(\frac{2}{\sin\left(x-\theta_n\right)}\right)^2
  \rightarrow\infty\text{.}
\end{equation*}

We define the set $S(x)$
\begin{equation}
\label{eq:c_setS}
S(x) = \{n:|x-\theta_n|\leq |a_n| = n^{-\gamma} 2\pi\}\text{.}
\end{equation}
Each $n$ is an element of $S(x)$ if $x$ is ``close to $\theta_n$''. Note
that $\theta_n = 2\pi \{ \alpha_n / 2\pi \hbar \}$, where $\{.\}$ is the
fractional part, not ``set'' and $\alpha_n$ are the eigenvalues of the base
Hamiltonian $H_0$.

Given that $\sin x\leq x$ for all $x\geq 0$, a lower bound for $B^{-1}(x)$
is obtained,
\begin{equation}
\label{eq:c_Bestimate}
  \begin{split}
  B^{-1}(x) &\geq \sum_{n=0}^\infty |a_n|^2
  \left(\frac{2}{x-\theta_n}\right)^2 \\
  &\geq \sum_{n\in S(x)} \frac{4|a_n|^2}{(x-\theta_n)^2}
  \geq 4 \#S(x)\text{.}
\end{split}
\end{equation}
Each $n\in S(x)$ gives a contribution to the sum of greater than one
as $|a_n| / |x-\theta_n | \geq 1$. By only considering $\#S(x)$, we simply
count a ``$1$'' each time.

The results on discrepancy of sequences are now used, with the
sequence $\omega_\text{HO} = (\theta_n / 2\pi)$. Note that
each element of the sequence $\omega_\text{HO}$ is in $[0,1)$.

Consider the interval, defined for every $x\in (0,2\pi)$ and centred
around $x/2\pi$,
\begin{equation}
\label{eq:c_intervalJ}
J_N(x) = \left[ \frac{x}{2\pi} - N^{-\gamma},
  \frac{x}{2\pi} + N^{-\gamma} \right]\text{.}
\end{equation}
For large enough $N$, $J_N(x) \subset [0,1)$. The size of the interval is
$2N^{-\gamma}$. Using this particular subset and noting that the definition
of discrepancy \refeq{eq:c_discrepancy} involves taking the supremum over
all subsets of $[0,1)$, Combescure obtains
\begin{equation*}
\left| N^{-1}A(J_N(x),N) - 2N^{-\gamma} \right| \leq D_N(\omega_\text{HO})
\text{.}
\end{equation*}
Multiplying through by $N$ gives
\begin{equation}
\label{eq:c_inequality}
\left| A(J_N(x),N) - 2N^{1-\gamma} \right| \leq ND_N(\omega_\text{HO})
\text{.}
\end{equation}
As $|\psi\rangle \notin l_1(H_0)$
\begin{equation*}
\sum |a_n| \rightarrow\infty
\end{equation*}
and thus
\begin{equation*}
1/2 < \gamma \leq 1
\end{equation*}
from simple convergence arguments. Therefore, $N^{1-\gamma}$ grows at a
rate\footnote{Interestingly, it can in fact not grow
at all ($\gamma = 1$) which is a subtle point seemingly missed by 
Combescure and others. The rank-$1$ projection operator from the vector
$|\psi\rangle$ constructed with $\gamma=1$ is not shown to lead to the
emergence of a continuous spectrum. Therefore, the statement that
$|\psi\rangle \notin l_1(H_0)$ implies $|\psi\rangle \in \mathcal{H}_\cont$
is not in fact proved to be true. There are vectors not in $l_1(H_0)$
that may not be in the continuous spectrum. In practice (numerical,
experimental work) this should not cause any trouble. It is clearly easy to
avoid $\gamma=1$.} less than $N^{1/2}$. At this stage, Combescure utilises
the theorems discussed above on the discrepancy of sequences. For the
eigenvalue sequence, $\alpha_n = n\hbar\omega$, of the
harmonic oscillator\footnote{Do not confuse $\omega$, the harmonic oscillator
frequency, with $\omega_\text{HO}$, the label for the sequence in $[0,1)$,
the discrepancy of which is being bounded.} the $j=1$ case of
Theorem~\ref{thrm:c_d_ub} applies which is exactly
(Theorem~3.2, \cite{Kuipers}). Combescure obtains the result\footnote{This is
not based on a conjecture as for $j=1$ a direct proof is possible, bypassing
Conjecture~\ref{conj:c_weyl}. See \cite{Kuipers}.}
\begin{equation*}
D_N(\omega_\text{HO}) = O(N^{-1(/\eta) + \epsilon})\text{.}
\end{equation*}
For the sequence $\omega_\text{HO}$, $\beta = \omega/2\pi$. If $\beta$ is an
irrational of constant type ($\eta = 1$), the strongest type of irrational,
then
\begin{equation*}
ND_N(\omega_\text{HO}) = O(N^\epsilon)\text{.}
\end{equation*}
As the right-hand side of \refeq{eq:c_inequality} can be made to grow
arbitrarily slowly, we conclude that the left-hand side must grow slowly
too. Thus, to cancel out
the growth of $2N^{1-\gamma}$, $A(J_N(x),N)$ must grow at a
rate arbitrarily close to that of $2N^{1-\gamma}$. We see that
\begin{equation*}
A(J_N(x),N)\rightarrow\infty
\end{equation*}
as $N\rightarrow\infty$. It is now a simple observation \cite{Combescure90}
that this implies that $\#S(x)\rightarrow\infty$ and thus
$B^{-1}(x)\rightarrow\infty$. Thus, $e^{ix}$ is in
the continuous spectral subspace of the Floquet operator $V$.

The importance of the eigenvalue sequence is seen in that if we cannot limit
the right-hand side of \refeq{eq:c_inequality}, then we cannot place a lower
limit on $A(J_N(x),N)$ and thus we cannot conclude that
$B^{-1}(x)\rightarrow\infty$. Two barriers to limiting the right-hand side
of this
equation exist---$j$ and $\eta$. If, still in the harmonic oscillator case,
we wished for $\beta = \omega/2\pi$ to only be of a weaker type, say
$\eta=2$, we would no longer be able to conclude that
$B^{-1}\rightarrow\infty$. The right-hand side would grow like
$N^{(1/2) + \epsilon}$,
which is always faster than $2N^{1-\gamma}$ for $1/2<\gamma\leq 1$ which
grows at a rate of $N^{(1/2) - \epsilon}$. Thus, no suitable lower limit for
$A(J_N(x),N)$ can be found. Similarly, if the eigenvalue sequence is
generalised (Combescure's conjecture) then we run into trouble. For $j=2$,
the lowest possible growth rate for the right-hand side we can obtain, taking
Conjecture~\ref{conj:c_weyl} as true, applying Theorem~\ref{thrm:c_d_ub} and
noting Theorem~\ref{thrm:c_d_lb} which says we cannot do any better, is
, once again, $N^{(1/2) + \epsilon}$. For larger $j$, the situation only
gets worse.

Given these seemingly significant problems, the natural question to ask is:
``How does one get around this problem?''. The answer is provided in the
work of Bourget \cite{Bourget02.1}. Bourget proves a weaker theorem than
Combescure's conjecture. Where the same requirement on $|\psi\rangle$ is
kept in \cite{Combescure90}, that it be in $l_1(H_0)$, Bourget has a $j$
dependent requirement. Essentially, for increasing $j$ the $a_n$ terms used
to construct $|\psi\rangle$ must decrease more slowly with $n$. See
Bourget's work for the exact requirement, which depends on the best
estimates available for Weyl sums discussed earlier and thus is a
non-trivial function of $j$.

The key insight in obtaining the proof is to modify the set $S(x)$
(equation \refeq{eq:c_setS}) and the corresponding interval $J_N(x)$
(equation \refeq{eq:c_intervalJ}) that we consider. Importantly, they
become $j$ dependent. Bourget reduces the shrinking rate of the set
$J_N(x)$ as a function of $N$ just enough so as to allow the
weaker limits on the discrepancy to be good enough to force the right-hand
side of the equivalent to \refeq{eq:c_inequality} to be less than the
left-hand side, while
keeping strong enough control on terms in the new set $S(x)$ to still argue
that $B^{-1}\rightarrow\infty$.

Using the best available estimations on Weyl sums and plugging these into
the upper bound formulas for discrepancy (as discussed above when introducing
the work by Vinogradov), Bourget manages to provide a rigorous proof of
the existence of a continuous spectral component of the Floquet operator
(the essence of Combescure's conjecture) for $j\geq 3$ \cite{Bourget02.1}
and $j=2$ \cite{Bourget05}. The proof is, unfortunately,
unavoidably clouded by the ``messy'' estimates available for Weyl sums and
thus, the essence of the proof is difficult to see. Here, we will revisit
the proof, but (utilising Conjecture~\ref{conj:c_weyl}) apply
Theorem~\ref{thrm:c_d_ub} which says (using
$2\epsilon$, rather than $\epsilon$ for technical reasons), for all
$\epsilon> 0$
\begin{equation*}
D_N(\omega) = O\left(N^{-1/(\eta j) + 2\epsilon}\right)\text{.}
\end{equation*}
With this very clean estimate, it is far easier to see how Bourget's work
provides a proof that a continuous spectral component of the
Floquet operator exists. It covers all $j\geq2$. We highlight the fact that
a solution to Vinogradov's conjecture provides an elegant solution to
Combescure's physics conjecture. The $j$-dependence of the $a_n$s used to
construct $|\psi\rangle$ is straightforward.

\begin{thrm}
\label{thrm:c_allj}
Assume Conjecture~\ref{conj:c_weyl} is true and thus
Theorem~\ref{thrm:c_d_ub} follows. Assume $\beta$ is irrational and of
type $\eta$. Then for all positive integers, $j$, the Floquet operator,
$V$, has $\sigma_\cont(V) \neq \emptyset$ if
$1/2 < \gamma < 1/2 + 1/(2 \eta j)$.
\end{thrm}

\emph{Proof.} (\ref{thrm:c_allj})
The proof relies upon the techniques utilised by Bourget. In essence, 
we simply increase the size of the interval (equation
\refeq{eq:c_intervalJ}) from
$2N^{-\gamma}$ to $2N^{2((1/2)-\gamma)}(\log N)^{-1/2}$. The important change 
is the first factor. The $\log N$ term is essential for
technical reasons, but has a negligibly small effect on the shrinkage rate
of the interval for large $N$. As $\log N / N^{4\delta} \rightarrow 0$ as
$N\rightarrow\infty$ for all $\delta>0$, for $N$ large enough we have
\begin{equation*}
2N^{2((1/2)-\gamma)}(\log N)^{-1/2} > 2N^{2((1/2) - \gamma - \delta)}\text{.}
\end{equation*}
Using this underestimate for the size of the interval, we easily obtain
the equivalent of \refeq{eq:c_inequality},
\begin{equation*}
\left| A(J_N(x),N) - 2N^{2(1 - \gamma - \delta)} \right| \leq ND_N(\omega_j)
\text{,}
\end{equation*}
for the sequence $\omega_j = (n^j\beta)$. Now, using
Theorem~\ref{thrm:c_d_ub}, it is evident that to ensure
$A(J_N(x),N)\rightarrow\infty$, we must have
\begin{equation*}
2(1 - \gamma - \delta) > 1-1/(\eta j) + 2\epsilon\text{,}
\end{equation*}
or
\begin{equation*}
\gamma < 1/2  + 1/(2\eta j) - \epsilon - \delta \text{.}
\end{equation*}
The condition
\begin{equation}
\label{eq:c_gengamma}
1/2 < \gamma < 1/2 + 1/(2\eta j)\text{,}
\end{equation}
where the ``$<$'' sign has absorbed the arbitrarily small numbers
$\epsilon$ and $\delta$, must be satisfied to force 
$A(J_N(x),N)\rightarrow\infty$.

Finally, we must show that $B^{-1}(x)\rightarrow\infty$ when this
larger interval is used. Corresponding to the new interval $J_N(x)$, we
introduce the new set $S(x)$,
\begin{equation*}
S(x) = \left\{n:|x-\theta_n|
  \leq 2\pi N^{2((1/2)-\gamma)}\log N^{-1/2}\right\}\text{.}
\end{equation*}
The estimate \refeq{eq:c_Bestimate} is the same, except with the new set
$S(x)$, which no longer has all terms greater than unity. Thus, it is not
enough to simply count the number of terms in $S(x)$. A more subtle
estimate is required. Replacing the numerator, $|a_n|$, with something
smaller, $N^{-\gamma}$, and the denominator, $(x-\theta_n)$, with something
larger, $2\pi N^{2((1/2)-\gamma)}\log N ^{-1/2}$, we obtain
\begin{equation*}
B^{-1}(x) \geq \frac{1}{\pi^2} \sum_{n\in S(x)}
  \frac{\log N}{N^{2(1-\gamma)}}\text{,}
\end{equation*}
which is essentially the estimate Bourget obtains. The estimate contained
therein (Lemma~3.5 in \cite{Bourget02.1}) then shows that
$B^{-1}(x)\rightarrow\infty$ and the argument is complete.
\proofend

Examining \refeq{eq:c_gengamma} we note that for $j=1$ (for $\eta=1$) we
recover the simple result of Combescure. For all $j\geq 2$, we have a stronger
($j$-dependent) condition on $|\psi\rangle$ than simply
$|\psi\rangle \notin l_1(H_0)$. This complication is the main weakening
of Combescure's conjecture that Bourget and we have been forced to make.
Note that the restriction on $\gamma$ takes into account the end point
subtleties referred to in the preceding discussions.

We have replaced the requirement that $|\psi\rangle\notin l_1(H_0)$
(i.e.,\ $1/2 < \gamma \leq 1$) with the $j$ dependent requirement
$1/2 < \gamma < 1/2 + 1/(2j)$. In Bourget's work, the requirement is
stronger---directly related to the replacement of the known limits on Weyl
sums (in terms of $\rho$ in the earlier sections) with the ``best
possible'' estimate from our Conjecture~\ref{conj:c_weyl} of
$(1/j)-\epsilon$.

\subsection{Summary}

Reliance on Conjecture~\ref{conj:c_weyl} and the result of
Theorem~\ref{thrm:c_d_ub} derived from it has allowed us to discuss
Bourget's proof without the complications of the messy estimations on Weyl
sums.  Bourget's proof is also $n$-dependent ($m$ in his work) while ours
is $n$-independent. This simplified discussion highlights the key aspects
of Bourget's proof, both for $j\geq3$ \cite{Bourget02.1} and $j=2$
\cite{Bourget05}. It has also shown that the emergence of a continuous
spectral component of the Floquet operator is solved by Vinogradov's
conjecture. A proof of Vinogradov's conjecture is no longer just of
mathematical interest.  It has a direct mathematical physics consequence.

Finally, note that the rank-$N$ equivalent of this work follows in the
same way as presented for the harmonic oscillator case in
\refsec{sec:c_comb2}, providing a complete rank-$N$ generalisation of the
work of Combescure \cite{Combescure90}.


\section{\label{sec:c_milek}Generalising the results of Milek and Seba}

Having established that the continuous subspace of $\mathcal{H}$,
$\mathcal{H}_\cont$ is not empty, we wish to characterise it---by
identifying the singular and absolutely continuous components. Here, we
extend the result of Milek and Seba to rank-$N$ perturbations.

\begin{thrm}
\label{thrm:c_ac_empty}
Assume $H(t)$ is given by \refeq{eq:c_hamiltonian} and that
\refeq{eq:c_measure} applies. Assume that $B_k^{-1}(x)\rightarrow\infty$
and thus $\mathcal{H}_\cont \neq \emptyset$.
Then $\mathcal{H}_\ac = \emptyset$ and thus $\mathcal{H}_\cs$ is
not-empty. The Floquet operator, $V$, has a non-empty singular
continuous spectrum.
\end{thrm}

\emph{Proof.} (\ref{thrm:c_ac_empty})
As shown in the proof of (Theorem~II.4a, \cite{McCaw05.1}) and easily
calculated, the Floquet operator can be written in the form
\begin{equation*}
	V = U+\sum_{k=1}^N R_k\text{,}
\end{equation*}
where
\begin{equation}
\label{eq:c_r_k}
R_k = \left(e^{i\lambda_k/\hbar} - 1\right)|\psi_k\rangle
  \langle\psi_k|U\text{.}
\end{equation}

We can now use either (Theorem~5, \cite{Howland79}) or
(Theorem~1, \cite{Birman}). The theorem from the paper of Birman and Krein is
more direct, so we use it here. It states that if we have two unitary
operators, $U$ and $V$, that differ by a trace class operator, then the wave
operators
\begin{equation*}
\Omega_\pm = \text{s-}\underset{\nu\rightarrow\pm\infty}{\lim}
	V^\nu U^{-\nu} P_\ac(U)
\end{equation*}
exist and their range is the absolutely continuous subspace of $V$,
\begin{equation}
\label{eq:c_birman}
R(\Omega_\pm) = \mathcal{H}_\ac(V)\text{.}
\end{equation}

We must show that the difference $V-U$ is finite. With the notation in
\cite{McCaw05.1}, where the perturbation $W$ is given by
$A^*A$ and
\begin{equation*}
A = |\psi\rangle\langle\psi|\text{,}
\end{equation*}
with
\begin{equation*}
|\psi\rangle = \sum_n a_n \phi_n\text{,}
\end{equation*}
we obtain
\begin{equation*}
\begin{split}
\tr A^*A = \tr A &= \sum_l \langle\phi_l|A|\phi_l\rangle \\
  &= \sum_{l,m,n} \langle\phi_l | a_n | \phi_n\rangle
     \langle\phi_m | a_m^* | \phi_l\rangle \\
  &= \sum_{l,m,n} a_n a_m^* \delta_{ln}\delta_{ml} \\
  &= \sum_l |a_l|^2 = 1
\end{split}
\end{equation*}
as $|\psi\rangle\in l_2(H_0)$ and is normalised. The perturbation to the
Hamiltonian is trace class. The difference in unitary operators, $U$ and
$V$, is also trace class. By the triangle inequality for norms,
\begin{equation*}
\lpar R_k \rpar_{\text{Tr}} \leq
\lpar \left(e^{i\lambda_k/\hbar}-1\right)\rpar\lpar|\psi_k\rangle
\langle\psi_k|\rpar_{\text{Tr}} \lpar U \rpar_{\text{Tr}}\text{.}
\end{equation*}
As $\lpar U \rpar_{\text{Tr}}=1$,
\begin{equation*}
\begin{split}
\tr \left( \sum_{k=1}^N R_k \right)
  &\leq \sum_k \lpar(e^{i\lambda_k/\hbar}-1\rpar
    \sum_{l,m,n} \langle\phi_l | (a_k)_n | \phi_n \rangle
    \langle\phi_m | (a_k)_m^* | \phi_l \rangle \\
  &= \sum_k \left | e^{i\lambda_k\hbar}-1 \right| \\
  &= \sum_k \sqrt{2(1-\cos \lambda_k/\hbar)}\text{.}
\end{split}
\end{equation*}
Armed with a trace-class perturbation, we conclude that the wave operators
exist. The existence of the operators $\Omega_\pm$ means that they are
defined for all vectors in the Hilbert Space $\mathcal{H}$. Note (equation
\refeq{eq:c_birman}) that the subspace $\mathcal{H}_\ac(V)$ is
equal to the range of these operators. However, $P_\ac(U)$ gives
zero when acting on any state in $\mathcal{H}$ because $U$ is pure point.
Thus, $\mathcal{H}_\ac(V)$ is empty. As $\mathcal{H}_\cont$
is not empty, $\mathcal{H}_\cs$ must be non-empty, and we have
proved that a singular continuous subspace of the Floquet operator $V$
exists.\proofend

The key assumption in Theorem~\ref{thrm:c_ac_empty} is that
$B_k^{-1}(x)\rightarrow\infty$. This is certainly true for $j=1$ if
$|\psi_k\rangle \neq l_1(H_0)$. For $j\geq 2$, Bourget
\cite{Bourget02.1,Bourget05} showed that
one can construct vectors $|\psi_k\rangle$ for which
$B_k^{-1}(x)\rightarrow\infty$. The results were discussed in
detail in \refsec{sec:c_genspec}. We have shown, in
Conjecture~\ref{thrm:c_allj}, that if Conjecture~\ref{conj:c_weyl} is
true then this result may be improved---the requirements on the states
$|\psi_k\rangle$ are less restrictive. The result was also extended to
rank-$N$ perturbations.

\subsection{Discussion}

Milek and Seba make a number of incorrect statements in obtaining this
result for the rank-$1$ case. Firstly, they state that
the operator\footnote{As we are dealing with the rank-$1$ case, the
subscript $k$ may be dropped from \refeq{eq:c_r_k}.}
$R = \left[\exp\left(i\lambda/\hbar\right)-1\right]|\psi\rangle\langle\psi|U$
is rank-$1$ which it is not---the presence of the unitary
operator $U$ stops $R$ from being rank-$1$. This is not, however, important.
The applicability of the theorems in \cite{Howland79,Birman} does not rely
upon the rank of the operator $R$, but upon it being of trace-class.
Secondly, they claim that the existence of the wave operators implies that
\begin{equation}
\label{eq:c_ac_spec}
\sigma_\ac(V) \subset \sigma_\ac(U)\text{.}
\end{equation}
This is, again, not true. Given that $\sigma_\ac(U)$ is
empty, it is indeed possible to conclude that $\sigma_\ac(V)$ is
empty, as discussed above, but the relation \refeq{eq:c_ac_spec} does not
follow. Consider the situation where $\sigma_\cont(U)$ is not
empty. Then there is a set of vectors in $\mathcal{H}$ which are continuous
for $U$. These vectors form the domain for the operator $V^\nu$ in the wave
operators. The action with $V^\nu$ does not however keep us in the subspace
$\mathcal{H}_\cont(U)$ as the space we get to (the range for
$V^\nu$) is only invariant for $\mathcal{H}_\cont(V)$, not
$\mathcal{H}_\cont(U)$. Thus, we may obtain a vector, necessarily
in $\mathcal{H}_\cont(V)$ due to invariance, but possibly in
$\mathcal{H}_\sing(U)$, and thus, we cannot conclude that
$\sigma_\ac(V) \subset \sigma_\ac(U)$.
These two points discussed do not make the final results of Milek and Seba
wrong, but ``only'' the proofs.

Of greatest concern is the use of Lemma~6 in \cite{Combescure90} without
justification. Milek and Seba have assumed that Combescure's conjecture is
true. It has taken fifteen years, and a significant amount of work by
Bourget, for that to be shown to be the case. We have shown that the
conjecture is directly linked to a long standing number-theoretic
conjecture. The work has also been extended to cover rank-$N$
perturbations.


\section{\label{sec:c_summary}Summary}

We have generalised the work of both Combescure \cite{Combescure90} and
Milek and Seba \cite{Milek90.1} from rank-$1$ to rank-$N$. We have also
discussed in detail Combescure's conjecture, our work on estimations of
discrepancy and the demonstration by Bourget \cite{Bourget02.1} that a
continuous spectral component of the Floquet operator does exist for
certain constructions of $|\psi\rangle$. This covers the essential aim of
Combescure's conjecture on the existence of a continuous spectral
component. A clear view of the essence of Bourget's proof has been
provided by taking a reasonable number-theoretic conjecture to be true.
With this clear view, the work of Bourget becomes more accessible. A
resolution to Vinogradov's conjecture would have direct implications
in mathematical physics.

An in depth critical analysis of the work of Milek and Seba was also
undertaken; we highlighted a number of misconceptions in the
work. A proof of Vinogradov's conjecture, allowing our work to provide an
elegant solution to Combescure's conjecture, remains desirable.


\addcontentsline{toc}{section}{Acknowledgements}
\section*{\label{sec:c_ack}Acknowledgements}

This work was supported by the Australian Research Council. We thank
Olivier Bourget for some useful discussions. One of the
authors, J.~McCaw, also wishes to acknowledge the financial support
provided by the David Hay Memorial Fund.


\nocite{p_Ames}

\addcontentsline{toc}{section}{References}
\bibliographystyle{hplain}
\bibliography{bibliography}

\end{document}